# Deep learning-based phase control method for coherent beam combining and its application in generating orbital angular momentum beams

Tianyue Hou,[1,†] Yi An,[1,†] Qi Chang,[1,†] Pengfei Ma,[1,3] Jun Li,[1] Liangjin Huang,[1] Dong Zhi,[2] Jian Wu,[1] Rongtao Su,[1] Yanxing Ma,[1] and Pu Zhou[1,4]

[1]College of Advanced Interdisciplinary Studies, National University of Defense Technology, Changsha 410073, China
[2]Hypervelocity Aerodynamics Institute, China Aerodynamics Research and Development Center, Mianyang 621000, China
[3]shandapengfei@126.com
[4]zhoupu203@163.com

**We incorporate deep learning (DL) into coherent beam combining (CBC) systems for the first time, to the best of our knowledge. Using a well-trained convolutional neural network DL model, the phase error in CBC systems could be accurately estimated and preliminarily compensated. Then, the residual phase error is further compensated by stochastic parallel gradient descent (SPGD) algorithms. The two-stage phase control strategy combined with DL and SPGD algorithms is validated to be a feasible and promising technique to alleviate the long-standing problem that the phase control bandwidth decreases as the number of array elements expands. Further investigation denotes that the proposed phase control technique could be employed to generate orbital angular momentum (OAM) beams with different orders by distinguishing the OAM beams of conjugated phase distributions.**

Coherent beam combining (CBC) of fiber lasers has great potential in breaking through the power limitation of a single laser beam while maintaining good beam quality, which has been widely studied during the past decades [1-7]. In high-power CBC systems, especially operation with serious thermal and experimental fluctuations, dynamic phase noise always occurs and affects the performance of the combined beam. To eliminate the influence of dynamic phase noise and achieve constructive interference in the far field, the relative phase of each array element should be accurately controlled. For actively phase-locked arrays, several approaches have been proposed, including heterodyne detection [8], multi-dithering and single frequency dithering technique [9-13], interferometric technique [14], phase-intensity mapping [15] and stochastic parallel gradient descent (SPGD) algorithm [3,16]. Along with simultaneous increase of combined numbers and output power, the control bandwidth of the phase controller would be a serious issue that should be considered carefully [17].

To further improve the control bandwidth, a fast and accurate phase extraction method is highly required. As a result of excellent real-time performance, deep learning (DL) and artificial intelligence algorithms may offer a robust route to further improve the phase control speed in CBC system. In fact, this new technique has been successfully applied to many optical research fields, such as mode-locked lasers, optical microscopy and laser mode decomposition [18-21]. In order to incorporate it into CBC system, there is an intuitive idea that constructing a DL network for controlling the relationship between the intensity profile of the combined beam and the relative phases of array elements. In theory, by analyzing the real-time collected intensity profile based on the network, the phase error of the beam array could be estimated and compensated. However, one significant difficulty encountered is that there is no one-to-one correspondence between the far field intensity profile of the combined beam and the relative phases of array elements, thus the DL network would lose its effectiveness due to data collision. Quite recently, the concept of extracting cost functions at the non-focal-plane has been proposed by our group [22]. Drawing on this concept, the primary difficulty by incorporating DL method into CBC technology could expect to be solved.

In this Letter, we present a DL-based, two-stage phase control method for conventional coherent beam combining and orbital angular momentum (OAM) beams generation. To avoid the data collision mentioned above, the non-focal-plane intensity profiles of the combined beam are used as training samples. We construct and train a convolutional neural network (CNN) for real-time estimating the relative phases of the array. After preliminarily compensate the estimated phase error, the residual phase error is further compensated by SPGD algorithms to achieve more accurate phase locking. Our simulation results demonstrate the feasibility of the two-stage phase control technique, which combines the advantages of DL and SPGD algorithms for improving the control bandwidth. Further, the feasibility of the proposed two-stage phase control strategy for generating high-power, mode-switchable OAM beams is validated.

Figure 1 shows the experimental configuration to implement the DL-based, two-stage phase control method. The linearly polarized seed laser (SL) is amplified by a pre-amplifier (PA), and it is then split into multiple channels by a fiber splitter (FS). The laser beam of each channel pass through a fiber phase modulator (FPM) and cascaded fiber amplifiers (FAs) for power scaling. Subsequently, the laser beams are emitted through a collimator array and propagate in free space. The collimated beam array is split into two parts by a high reflective mirror (HRM). The transmission part propagates through another HRM and a focus lens (FL), and then is sampled by a beam splitter (BS) for joint feeding the phase control system at the non-focal-plane and the focal plane. Specifically, our phase control scheme consists of two stages, i.e. (i) estimating and compensating the phase error by CNNs and (ii) further compensating the residual phase error by SPGD algorithms. The intensity profile of the combined beam collected by the CCD locates at the non-focal-plane is sent to a FPGA controller for performing the first stage, while the intensity profile collected at the focal plane is sent to the FPGA controller to perform the second stage. The FPGA controller carrying a well-trained CNN and SPGD algorithm performs the two stages continuously and applies control voltages to the FPMs to achieve phase locking.

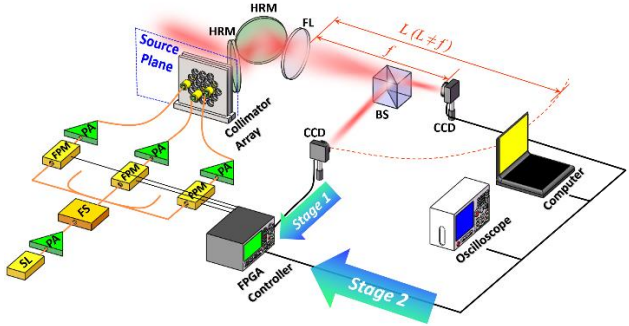

Fig. 1. Experimental configuration to implement DL-based, two-stage phase control method for CBC. (SL: seed laser; PA: pre-amplifier; FS: fiber splitter; FPM: fiber phase modulator; FAs: fiber amplifiers; HRM: high reflective mirror; FL: focus lens; BS: beam splitter.)

The electric field of a *N*-element linear polarized fundamental mode Gaussian beam array at the source plane can be expressed as follows:

$$E(\boldsymbol{\rho}, z=0) = \sum_{j=1}^{N} A_0 \exp\left[-\frac{(\boldsymbol{\rho}-\boldsymbol{\rho}_j)^2}{w_0^2}\right] circ\left(\frac{|\boldsymbol{\rho}-\boldsymbol{\rho}_j|}{d/2}\right) \exp(i\varphi_j),$$ (1)

where $A_0$, $w_0$, $\varphi_j$ and $d$ are the amplitude, waist width, initial phase and aperture diameter of the *j*th beamlet, respectively. $\boldsymbol{\rho}=x\hat{\boldsymbol{x}}+y\hat{\boldsymbol{y}}$, and $\boldsymbol{\rho}_j=x_j\hat{\boldsymbol{x}}+y_j\hat{\boldsymbol{y}}$ represents the position vector of the *j*th beamlet. Under the paraxial approximation, the intensity profile of the combined beam can be represented by Fourier transform form as

$$I(\boldsymbol{r}, z=L) = \left|\frac{e^{i\frac{k}{2L}r^2}}{i\lambda L} \mathscr{F}\left\{E(\boldsymbol{\rho}, z=0)e^{i\frac{k}{2}\left(\frac{1}{L}-\frac{1}{f}\right)\rho^2}\right\}\right|^2,$$ (2)

where $\boldsymbol{r}$ denotes the position vector at the receiver plane. $\lambda$, $f$ and $L$ account for the wavelength, focal length and propagating distance, respectively. $\mathscr{F}\{\cdot\}$ denotes the Fourier-transform operation. Note that the distance between the collimator array and the FL is far less than $f$ and $L$. The intensity profiles of the non-focal-plane ($L\neq f$) are input into the CNN for previous training and real-time phase error compensation.

The CNN performed in the first stage of our phase control scheme is modified from the VGG-16 model [23], as shown in Fig. 2. The model is modified based on the input and output of the network. Concretely, the filter size of the first convolutional layer of VGG model is changed from 3×3×3 to 3×3×1, as our input is a single intensity pattern image. ReLU function is chosen for nonlinear activation after each convolutional layer, followed by max pooling [23]. The Softmax function after the last fully-connected (FC) layer of the original VGG model is replaced by Sigmoid for our regression problem. The network learns to estimate relative phase from a single intensity pattern. The intensity patterns are generated by randomly changing the relative phase of each element as the samples for training. Then, due to the sigmoid function before the output, the *N-1* phase vector is linearly scaled to [0, 1] by dividing $2\pi$ as a label of the corresponding pattern.

In the training procedure, the input images are passed through the layers of the CNN and regressed into a *N-1* output vector. We define the loss of our network as mean-square error (MSE) between the output and the label vector. Then, the parameters of CNN are updated iteratively using back-propagated gradients based on the MSE loss. We train the samples on a desktop computer with an Intel Core i7-8700 CPU and GTX 1080 GPU.

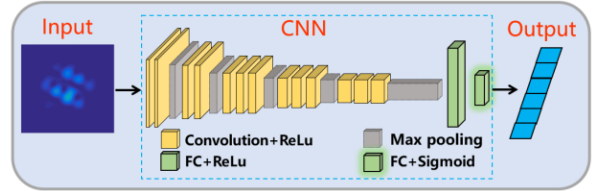

Fig. 2. Illustration of CNN.

When the network gets convergence, it can be utilized for estimate relative phase. Taking an intensity pattern image as input, the CNN output a *N-1* vector, from which the estimated relative phase can be obtained by multiplying $2\pi$.

In our previous work, we have indicated that the same far field intensity profile of a symmetrical beam array could correspond to different phase distributions in near field [22]. Here, we will explain how this problem affects the accuracy of the above DL network. Without loss of generality, a 7-element hexagonal array is taken as an example, shown in Fig. 3(a). The parameters of the array elements $w_0$, $d$, and $\lambda$ are assumed to be 23 *mm*, 10.24 *mm*, and 1.06 $\mu m$, respectively. The focal length is 20 *m*, and the non-focal-plane is 0.6*m* behind the focal plane.

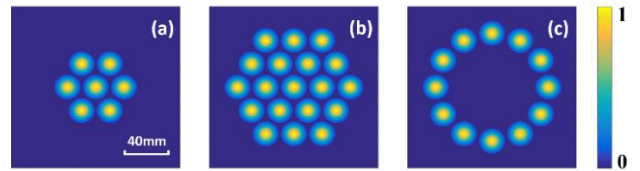

Fig. 3. Intensity profiles of the beam array consists of (a) 7 elements, (b) 19 elements, and (c) 12 elements.

Figure 4 exhibits the performances of the CNNs trained by intensity profiles of the non-focal-plane and the focal plane for the 7-element array. If the intensity profiles of the focal plane are used to train the CNN in advance and estimated the phase error in the first stage phase control, the far field energy concentration of the phase compensated combined beam [e.g., Figs. 4(b1)-4(b5)] is significantly lower than in the case of non-focal-plane [e.g., Figs. 4(c1)-4(c5)]. In other words, the CNN trained at the non-focal-plane could reflect the phase distribution of the beam array more accurately.

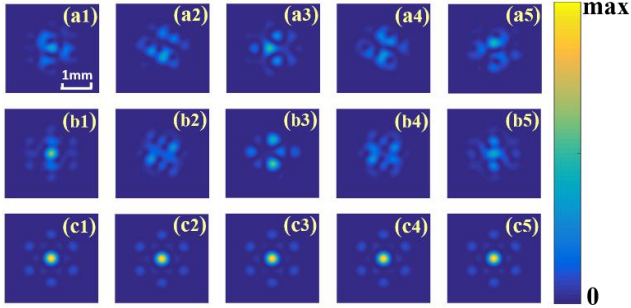

Fig. 4. Performances of the trained CNN in the first stage phase control. Far field intensity profiles (a1)-(a5) without phase error compensation, with phase error compensation using CNNs trained at (b1)-(b5) the focal plane and (c1)-(c5) the non-focal-plane.

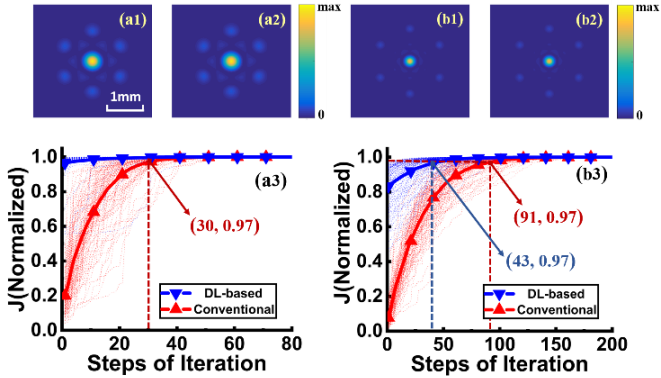

Fig. 5. (a1), (b1) Average far field intensity profiles of the 7-element and 19-element arrays after the first stage compensation. (a2), (b2) Average far field intensity profiles of the 7-element and 19-element arrays after the second stage compensation. (a3), (b3) Convergence curves of the cost functions of the 7-element and 19-element arrays. The radius of the bucket to estimate cost functions is $0.61\lambda D/f$, where $D$ represents the total diameter of the array. 100 times simulation has been involved for each case.

Based on the completion of the first stage, SPGD algorithm is implemented to further compensate the residual phase error. We use the power-in-the-bucket (PIB), which describes the energy encircled in an on-axis circular area at the receiver plane, as the cost function J of SPGD algorithm [3,24,25]. In general, we investigate the performances of our phase control method implemented in coherent combining of 7-element and 19-element hexagonal arrays, as shown in Fig. 5. The conditions and parameters for the 7-element hexagonal array case remain same as in the above analysis. As for the 19-element hexagonal array case, the arrangement of the beam array is shown in Fig. 3(b), and the non-focal-plane is 0.4 $m$ behind the focal plane. The second stage of the phase control method starts from the pre-compensated far field intensity profiles [e.g., Figs. 5 (a1) and 5(b1)], which already have central mainlobes with energy concentration by DL phase controlling. Hence, compared to conventional SPGD algorithm phase control method, DL-based, two-stage phase control method has fewer convergence steps, which means that the phase control bandwidth is improved, as shown in Figs. 5(a3) and 5(b3). As for the 7-element array, the normalized cost function after the first stage compensation is 0.97, and the DL-based phase control method saves 30 convergence steps on average. As for the 19-element array, the normalized cost function after the first stage compensation is 0.83. We define the criterion for convergence as the normalized cost function reaching 0.97, and the DL-based method saves 48 convergence steps on average. Actually, when the convergence criterion takes different values from 0.83 to 0.97, the average convergence steps saved by the DL-based method change little. With the array elements expanding from 7 to 19, despite that the residual phase error after the first stage compensation of the DL-based method increases, the saved convergence steps of the DL-based method also correspondingly increase, which could efficiently compensate the decrease of control bandwidth along with elements expanding. Furthermore, it could be inferred that the proposed phase control method has the potential to save more convergence steps as the number of array elements increases.

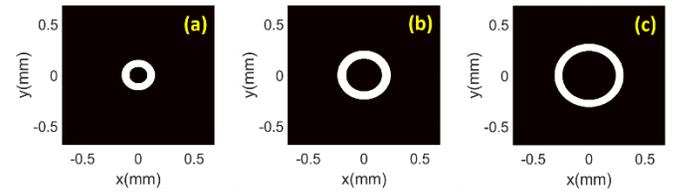

Fig. 6. Far field ring-shaped area for estimating the cost functions to generate (a) OAM ±1, (b) OAM ±2, and (c) OAM ±3 beams.

The above analysis concerns about eliminating dynamic phase noise in conventional CBC systems to achieve constructive interference in far field. In the following, we will show that the phase control method could also be used in locking the relative phase of each element to different value for generating complex light fields. As an example, the generation of OAM beams based on a 12-element ring-shaped array is investigated. At the source plane, the vortex phase structure is approximately composed by discontinuous piston phases of elements, and the OAM beams can be formed in the far field [26-28]. Here, we attach the piston phase to each element between the two stages of phase control, which can be realized by FPGA and FPMs. The "buckets" for estimating the cost functions to generate OAM±1, OAM±2, and OAM±3 beams are shown in Figs. 6(a)-6(c), respectively.

Figure 7 shows the results of OAM beams generation based on our phase control method. The piston phase attached to each element between the two stages of phase control for generating OAM -1, 1, +2, and +3 beams is shown in Figs. 7(a1)-7(a4). The convergence curves of the cost functions and the average intensity and phase distributions over 100 times are depicted in Figs. 7(b1)-7(b4). The non-focal-plane for extracting the intensity profiles sent to the CNN is 0.3 $m$ behind the focal plane. The results indicate that our method not only has advantages in convergence steps reduction, but can avoid the confusion caused by conjugated phase distributions as well. Specifically, OAM+1 and OAM-1 beams have the same far field intensity profiles, so when the cost function takes its maximum value, it corresponds to two conjugated phase distributions. Before the second stage of our method, the phase error has been preliminarily compensated by the CNN and the piston phases have been attached, thus the combined beam is quite similar with the expected OAM beam. Therefore, our method guarantees the avoidance of the local optimum and ensures that the OAM order of the combined beam converge to a definite value. This advantage in generating high-power, mode-switchable OAM beams is unique to the DL-based, two-stage phase control method, which is extremely difficult to be achieved with conventional phase control methods. Although we have only presented the typical case of a 12-element array, the phase control method has been validated to be highly available in coherent combining of different arrays for generating OAM beams with various orders.

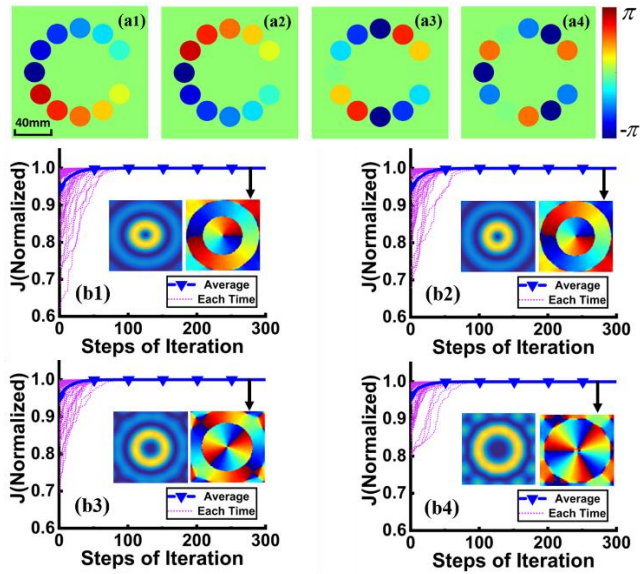

Fig. 7. Generation of OAM beams. (a1)-(a4) Piston phase attached to each element for generating OAM -1,+1,+2, and+3 beams. (b1)-(b4) Convergence curves of the cost functions for generating OAM -1,+1,+2, and+3 beams. 100 times simulation has been involved for each case. The inset figures show the average intensity (left) and phase (right) distributions of the generated OAM beams.

In conclusion, we present a DL-based, two-stage phase control method for CBC systems. Compared to conventional SPGD algorithm phase control method, our method can efficiently reduce the convergence steps. Besides, by modulating the cost function and attaching piston phases, high-power, mode-switchable OAM beams can be generated from a CBC system. We believe that the proposed phase control method has great potential in improving the control bandwidth of high-power CBC systems with large number of beamlets, and achieving the generation and flexible switch of structured light with complex intensity and phase distributions.

**Funding.** National Natural Science Foundations of China (NSFC) (61705264, 61705265)

**Acknowledgment**. At least a portion of the technology which is discussed in this paper is the subject of one or more pending patent applications.

†These authors contributed equally to this Letter.